# The study of contact properties in edge-contacted graphene-aluminum Josephson junctions

Zhujun Huang[1], Neda Lotfizadeh[2], Bassel H. Elfeky[2], Kim Kisslinger[3], Edoardo Cuniberto[1], Peng Yu[2], Mehdi Hatefipour[2], Takashi Taniguchi[3], Kenji Watanabe[4], Javad Shabani[2], Davood Shahrjerdi[1,6]

[1] Electrical and Computer Engineering, New York University, Brooklyn, NY 11201
[2] Department of Physics, New York University, New York, NY 10003
[3] Center for Functional Nanomaterials, Brookhaven National Laboratory, Upton, NY 11973, USA
[4] International Center for Materials Nanoarchitectonics, National Institute of Materials Science, 1-1 Namiki Tsukuba, Ibaraki 305-0044, Japan
[5] Research Center for Functional Materials, National Institute of Materials Science, 1-1 Namiki Tsukuba, Ibaraki 305-0044, Japan
[6] Center for Quantum Phenomena, Department of Physics, New York University, NY 10003

Corresponding email: jshabani@nyu.edu, davood@nyu.edu

**Transparent contact interfaces in superconductor-graphene hybrid systems are critical for realizing superconducting quantum applications. Here, we examine the effect of the edge-contact fabrication process on the transparency of the superconducting aluminum-graphene junction. We show significant improvement in the transparency of our superconductor-graphene junctions by promoting the chemical component of the edge contact etch process. Our results compare favorably with state-of-the-art graphene Josephson junctions. The findings of our study contribute to advancing the fabrication knowledge of edge-contacted superconductor-graphene junctions.**

Graphene, due to its Dirac-like energy band structure, is a popular material for studying the interplay between superconductivity and quantum transport [1-6]. Two technical advances in fabricating graphene devices underlie this progress. The first is the ability to achieve ballistic transport in graphene through encapsulation in hexagonal boron nitride (hBN) [7]. With this approach, the mean-free path ($L_{mfp}$) of hBN encapsulated graphene heterostructures (BGB) at cryogenic temperatures can exceed 20 $\mu$m [7-9]. The second advance is the adaptation of the edge contact process [7] for making superconducting contacts to BGB heterostructures [10]. The successful combination of these two features has enabled the studies of quantum Hall-based superconducting edge states [3-6] and new types of superconducting quantum circuits [11-15] in BGB Josephson junctions (JJ).

Recent studies have established the importance of the transparency at the superconductor-semiconductor interface for exploring proximity-induced superconductivity in the hybrid systems [16-18]. One figure-of-merit $I_cR_n$ is often used to quantify the contact transparency in JJ devices [19], where $I_c$ is the critical supercurrent and $R_n$ is the normal resistance of the junction. Theory predicts that graphene JJ with transparent contact should achieve $I_cR_n = 2.44\Delta/e$ in the short limit, where $\Delta$ is the superconducting gap [20].

Previous demonstrations of edge-contacted graphene JJ devices reported significantly smaller $I_cR_n/\Delta$ than theory [2, 10, 12-14, 21]. A possible reason for this observation is the short coherence length ($\xi$) of the superconductors (e.g., Nb, NbN, MoRe) in those studies, which is comparable to or smaller than the graphene channel length ($L_{ch}$). This design constraint prevents the operation in the short limit; thereby making it difficult to assess the extent of graphene-superconductor transparency using $I_cR_n/\Delta$.

A recent study reported a significant improvement in the normalized $I_cR_n$ for edge-contacted graphene JJ devices, establishing the prospects of achieving operation in the short limit [22]. A key to this demonstration was the fabrication of short-channel ($L_{ch}<$ 0.2 $\mu$m) graphene devices that were contacted by an aluminum (Al) superconductor, which satisfies a requirement for short limit operation, i.e., $L_{ch} \ll \xi$ (coherence length of aluminum). More importantly, the high $I_c$ (about 6 $\mu$A) of these short-channel graphene devices indicate high transparency at the superconductor-graphene interface. While edge-contacted graphene JJ has improved considerably, there is still significant room for enhancing their performance to reach the theoretical prediction. Studies of the structural, chemical, and electronic properties of the superconducting contacts to graphene are crucial for developing optimization strategies. Our work here is a step in this direction.

Here, we fabricated edge-contacted graphene devices with Al superconducting contacts as the test vehicle. Specifically, we examined the effect of the edge contact etch process on the performance of the resulting JJ devices. We observed that modifying this step can lead to significant enhancement in $I_cR_n/\Delta$ that outperforms JJ graphene devices with large $\Delta$ superconducting electrodes [2, 10, 12-14, 21], while matching the "second-to-champion" devices with Al superconducting electrodes [22]. Lastly, we examined the structural properties and elemental composition of the contacts to graphene in our BGB JJ devices using high-resolution transmission electron microscopy (HRTEM). This analysis revealed the unintentional incorporation of carbon (C) and oxygen (O) impurities at the contact interface with graphene.

In this work, we used poly(vinyl) alcohol (PVA)-assisted-graphene-exfoliation method to produce the monolayer graphene flakes [23]. A sub-10 nm PVA film was spin-coated onto Si substrates covered with $SiO_2$ and used as the adhesion promotion layer during the graphene exfoliation. The BGB heterostructures were constructed using the Quantum Material Press tool [24]. To achieve BGB heterostructures with atomically clean interfaces, we followed a stacking process, which allows the full removal of the polymeric contaminations across the entire dimensions of the heterostructures [8]. Before the fabrication of devices, we confirmed the monolayer graphene and the interface cleanliness using Raman spectroscopy [25] (see Supplementary Fig. S1).

Figure 1a shows the schematic illustration of a BGB JJ device, where graphene is edge-contacted by bilayer Ti/Al (10/30 nm) electrodes. In this structure, the silicon substrate functions as the global bottom gate. We observed that direct deposition of Al yields poor electrical contact to graphene. In contrast, we found experimentally that the incorporation of a 10-nm-thin Ti interlayer considerably improved the contact quality.

We employed a self-aligned process for fabricating the BGB JJ devices [10]. This fabrication process is useful for minimizing the polymer residues at the contact region, while relaxing the requirement of good alignment in the lithography step. We fabricated the graphene JJ devices by first defining a rectangle-shaped BGB mesa using a combination of e-beam lithography and an etching step. A subsequent lithography step defined the self-aligned metal contact regions, followed by reactive ion etch (RIE) of the BGB heterostructures that exposes the graphene edge for contacting to the metal electrodes. Lastly, Ti/Al metal electrodes were formed through sputtering and lift-off processes.

We fabricated two sets of device samples (Al-1 and Al-2) by varying the edge contact etch process. Specifically, we employed two different $CHF_3/O_2$ gas mixtures by adjusting the flow rates (40/4 sccm for Al-1 and 60/4 sccm for Al-2 samples). Other etch conditions in this step were kept unchanged. Our rationale for this experimental design is that increasing this gas ratio is known to promote the chemical etching [26]. We hypothesized that promoting the chemical component of the BGB etch can be favorable for enhancing the chemical activation of the graphene edge to yield stronger coupling to the metal electrode [27]. Figure 1b shows the optical image of Al-1 JJ devices configured in a transfer-length-measurement (TLM) structure with $L_{ch}$=0.5, 0.75, and 1 μm. We designed the JJ devices on Al-2 to include smaller $L_{ch}$ than Al-1, with $L_{ch}$=0.3, 0.4, and 0.6 μm. The channel widths of all devices are W=5 μm.

Our electrical characterization of the BGB JJ devices initially focused on studying carrier transport in graphene at room temperature (Figure 1) and low temperature of 1.5 K (Figure 2). These measurements were performed using standard low-frequency lock-in technique with an excitation current of 5-10 nA. The BGB JJ devices had a quasi-four-point structure, which eliminates the resistance contribution from the metal electrode leads. The current was injected from lead 1 to 2 and the voltage was measured between leads 3 and 4 (see Figure 1b). This configuration measures the sum of the graphene channel resistance and contact resistance.

Figures 1c-d show the room-temperature device resistance (R) as a function of the gate bias ($V_g$) for JJ devices on the Al-1 and Al-2 samples, respectively. In all devices, the total resistance exhibits an electron-hole asymmetry, where the hole branch has a higher resistance than the electron branch. This electron-hole asymmetry is attributed to the contact-induced electron doping [2, 10, 28]. This phenomenon yields the formation of p-n junctions at the contact region, creating an additional barrier for the hole carriers at the metal-graphene interfaces.

We employed TLM analysis to examine the contact resistance and carrier transport at room temperature. The width-normalized resistance can be written as a linear function of $L_{ch}$ as follows

$$RW = R_c W + R_{ch} W = R_c W + \frac{L_{ch}}{\sigma}$$

where $R_c$ is the total contact resistance, $R_{ch}$ is the graphene channel resistance and $\sigma$ is the graphene conductivity. By applying a linear fit to the TLM data (see Supplementary Fig. S2), we extracted the width-normalized $R_c$ (see Figures 1e and g). We also calculated the mean-free-path ($L_{mfp}$) by assuming diffusive transport in these devices. Figures 1f and h show $L_{mfp} = \frac{h}{2e} \frac{\sigma}{k_F}$ as a function of carrier density, where $k_F$ is the Fermi wavevector. This analysis revealed that Al-2 devices have a smaller $R_c$ than Al-1 devices at high carrier density. Furthermore, we calculated 0.5 $\mu$m<$L_{mfp}$<1 $\mu$m in Al-1 devices at high electron density, which is comparable to the device dimension (see Figure 1f). The resistance of the Al-2 devices at high electron density showed negligible channel length dependence, which yielded $L_{mfp} \gg 0.6$ $\mu$m (marked by the gray shading in Figure 1h). The TLM analysis at room temperature suggests that the Al-1 and Al-2 devices operate in the crossover regime between the diffusive and ballistic transport.

Next, we studied the low-temperature transport at T=1.5 K, which is above the superconducting critical temperature ($T_c$) of Al. Figures 2a-b show the gate voltage dependent plots of resistance for Al-1 and Al-2 devices, respectively. The data revealed negligible channel-length dependence of resistance away from the CNP, suggesting ballistic transport. Assuming ballistic transport in graphene, we investigated the metal-graphene contact properties. A ballistic conductor (i.e., $L_{mfp} > L_{ch}$) with no scattering centers has zero resistance. Therefore, the total resistance in a ballistic device can be attributed to the resistance in the immediate vicinity of the conductor-contact interfaces, which is known as the Sharvin resistance ($R_s$) [29-31]. This quantum-limited resistance is determined by the number of conducting modes (M) in the ballistic conductor. Thus, the graphene Sharvin resistance can be written as

$$R_s^{-1} = G_s = g_0 M = \frac{4e^2}{h} \text{int}(\frac{k_F W}{\pi})$$

where h is the Planck constant, $g_0$ is the conductance quantum, and the factor 4 comes from the spin and valley degeneracy in graphene. In Figures 2a-b, we plotted $R_s$ to allow comparison with the measured device resistances. This analysis revealed an observable difference between the theoretical and measured values, suggesting that additional factors should be considered for describing the device resistance.

In Landauer's formula, a finite transmission probability ($T_r$) is used to account for the difference in R and $R_s$ [32]. $T_r$ describes the averaged probability of electron transmission from one metal contact to the other. Thus, the overall conductance G can be written as $G=G_s T_r$. From this equation, we extracted $T_r$ as a function of the carrier density in graphene, as shown in Figures 2c-d. This analysis revealed that, at high electron density, Al-2 devices exhibit about 30% higher $T_r$ than the Al-1 devices ($T_r$~0.4 *versus* 0.3). While this analysis estimates the extent of the non-ideality in the ballistic transport of graphene devices, further studies are required to identify the exact sources of the finite $T_r$. For example, various non-idealities in the materials can contribute to the deviation of $T_r$ from unity, such as disorders at the contact interfaces (i.e., graphene edge or metal electrode), non-specular boundary scattering in graphene, and mismatch of conducting modes between metal and graphene [29, 31]. Nevertheless, this analysis revealed the noticeable role of the edge contact etch in our fabrication process in shaping the electronic properties of the contact.

Next, we evaluated the Josephson characteristics of the Al-1 and Al-2 devices. Figure 3 shows the representative Josephson dc measurements of the 0.4 μm device on the Al-2 device sample at 15 mK. The data show $I_c$ of 1.63 and 0.3 μA at electron and hole carrier density of $1.4\times10^{12}$ and $-5\times10^{11}$ cm$^{-2}$, respectively. These results confirm the gate tunability of the supercurrent. $I_c$ increases with the carrier density on both electron and hole branches. Moreover, the minimum $I_c$ (0.06 μA) occurs at the charge neutrality point (CNP). The differential resistance map in Figure 3b revealed the electron-hole asymmetry in $I_c$. For the same carrier density, the electron branch provides over 4 times higher $I_c$ than the hole branch. The considerably smaller $I_c$ on the hole branch is due to the contact-induced doping [2, 10, 28]. These results confirm the proximity-induced superconductivity of graphene. We have provided the results of the other devices on Al-1 and Al-2 samples in Supplementary Fig. S3.

Next, we examined the supercurrent interference pattern in the presence of vertical magnetic field ($B_z$). These measurements provide information about the field profile in the junction by analyzing the node periodicity of the supercurrent [33, 34]. Assuming a rectangular junction with uniform current distribution, the interference pattern follows

$$I_c(B_z)=I_c^0 \left| \frac{\sin \pi\Phi/\Phi_0}{\pi\Phi/\Phi_0} \right|$$

where $\Phi_0$ is the flux quantum, $\Phi$ is the out-of-plane flux in the junction region, $I_c^0$ is the zero-field critical current. In Figures 3c-e, we plot the measured Fraunhofer patterns at different carrier density of $1.4\times10^{12}$, $-5\times10^{11}$ cm$^{-2}$, and CNP. The field oscillation frequency is 0.43 mT, which gives $L_{ch}+2\lambda=0.93$ μm (hence $\lambda=0.26$ μm), where $\lambda$ is the penetration depth into the Al electrodes.

Furthermore, we used these datasets to reconstruct the current density distribution of the junction [35]. Figure 3f illustrates the critical current extracted from Figure 3c. In Figure 3g, we calculated current density profile, which shows nearly constant current density within the junction. This result implies a uniform contact quality and homogeneous transport properties in this BGB JJ device (see Supplementary Fig. S3 for measurements in other junctions).

The large $I_c$ of 1.63 μA in Figure 3a points to a good transparency at the Al-graphene junction. Therefore, we next examined the transparency of the superconducting Al-graphene junction by calculating $I_cR_n$. Figures 4a-b show the I-V curves of the Al-1 and Al-2 devices, measured at high electron density using a biasing current up to 10 μA. The use of high biasing current is an important consideration for accurate extraction of $R_n$ as it must be extracted in the regime where the junction becomes fully ohmic. From the slope of these I-V curves, we calculated $R_n \approx 90$ Ω for Al-1 devices at $5 \times 10^{11}$ cm$^{-2}$. For Al-2 devices, we obtained $R_n \approx 50$ Ω at $1.4 \times 10^{12}$ cm$^{-2}$. These extracted $R_n$ values are consistent with the transport measurements at 1.5 K. Furthermore, we used the cold branch I-V curves in Figures 4a-b to obtain $I_c$, providing $I_c$=0.6, 0.46 and 0.38 μA for Al-1 devices and $I_c$=1.7, 1.63 and 1.45 μA for Al-2 devices.

Figure 4c shows the summary of the $I_cR_n$ data for Al-1 and Al-2 devices normalized to Δ of the superconductor (notice the star symbols in this plot). An important consideration in obtaining $I_cR_n/\Delta$ for evaluating the contact transparency is to use the bulk superconducting gap of Al [17]. To do so, we measured $T_c$ of Al, which was 1.15 K and 0.9 K in Al-1 and Al-2 samples. Correspondingly, the bulk superconducting gap of Al in Al-1 and Al-2 samples were 175 and 136 μeV. Using this information, we calculated $I_cR_n/\Delta$ to be 0.3, 0.24, and 0.2 for Al-1 devices, 0.68, 0.6, and 0.54 of Al-2 devices. Comparing $I_cR_n/\Delta$ of the devices with similar $L_{ch}$ on Al-2 and Al-1 samples indicates almost 2 times improvement in the transparency of the superconducting Al-graphene junction. This analysis provides further evidence for the important role of the edge contact etch in our fabrication process.

To put these results in perspective, in Figure 4c, we compared $I_cR_n/\Delta$ of our devices with state-of-the-art counterparts that use large Δ superconductors [2, 10, 12-14, 21, 36]. For fair comparison in this plot, we normalized $I_cR_n$ to the bulk superconducting gap for all data. This summary plot indicates higher $I_cR_n/\Delta$ of our devices relative to their counterparts with similar lengths. We attribute the observed improvement, in part, to the longer coherence length of Al in our devices.

The measurements of $I_cR_n$ as a function of temperature can also provide information about the contact transparency [19]. Figure 4d shows the experimental $I_cR_n$ for the 0.5 and 0.3 μm devices on Al-1 and Al-2 samples. From the data, we recovered $I_cR_n = \alpha\Delta/e$ using the Kulik-Omelyanchuk relation [37] and found $\alpha$=0.22 and 0.43 for the 0.5 and 0.3 μm devices, respectively (see Supplementary Note 2). This analysis further confirms the improved junction transparency in the Al-2 devices.

Another important property of a JJ device is the induced superconducting gap $\Delta_{ind}$, which can be obtained from the multiple Andreev reflection (MAR) [17]. Figures 4e-f show the measured conductance of the 1 and 0.3 µm devices (on Al-1 and Al-2 samples) at different gate voltages. Several discernable conductance peaks can be observed in the conductance. The position of those peaks corresponds to the energy levels of $2\Delta_{ind}/N$, where N is an integer number. From Figures 4e-f, we extracted $\Delta_{ind}$=100 and 80 µeV for Al-1 and Al-2 devices. Our observation of a smaller $\Delta_{ind}$ than the bulk superconducting gap is consistent with the previous studies on Al devices [17]. This analysis emphasizes the importance of normalizing $I_cR_n$ to the bulk superconducting gap for assessing the junction transparency.

Lastly, we analyzed the structural properties and elemental composition of the metal-graphene junction using HRTEM and electron dispersive X-ray spectroscopy (EDS). Figure 5a shows the cross-sectional HRTEM image of a BGB JJ with $L_{ch}$=0.5 µm. Figure 5b shows the zoomed-in view of the same BGB JJ at the edge contact region. The HRTEM images confirmed the continuity of the Ti/Al metal stacks on the BGB sidewall. It also revealed the sharpness of the metal-BGB interface. The basal plane of the BGB layers is visible in Figure 5b, confirming the atomically clean interfaces between graphene and hBN layers. These results confirmed the excellent structural quality of the BGB stacks.

Figures 5c-g show the individual EDS maps of different elements obtained from the same region as Figure 5b. These results together with the overlapped EDS image in Figure 5h provided detailed information about the elemental composition of the metal-graphene junction. Figure 5h confirms the formation of the edge contact between graphene and the Ti layer. Curiously, this analysis also revealed the presence of elemental C and O atoms, which mainly overlap with the Ti interlayer. In our experiment, we did not investigate the mechanism of O and C incorporation and the nature of their interactions with Ti. However, we suspect that the incorporation of these impurities occurs during the Ti deposition due to its known high reactivity with C and O [38-40]. While we do not know how these impurities affect the junction transparency, we conjecture that they may be a limiting factor in improving the performance of our graphene JJ devices. Therefore, a future research direction is to investigate the effect of alternative interlayers that are less susceptible to the unintentional incorporation of C and O impurities.

In conclusion, the results reported here illustrate the noticeable role of the edge contact etch in our fabrication process on the transparency of the superconducting Al-graphene junction. Specifically, we found that promoting the chemical component of the etch process was beneficial in improving the device performance. With this modification, the normalized $I_cR_n$ of our devices surpassed that of state-of-the-art counterparts with large $\Delta$ (Refs. [2, 10, 12-14, 21, 36]), while matching the second-to-champion devices with Al (Ref. [21]; see Supplementary Fig. S4). Furthermore, our HRTEM study revealed the presence of C and O impurities in the Ti interlayer. We conjecture that further improvement of junction transparency may require strategies for achieving an impurity-free metal interlayer. Lastly, many studies have contributed to the persistent development of the edge contact fabrication process [7, 27, 41], and this work adds an advance in this direction for improving the superconductor-graphene junction.

**Acknowledgments**

D.S. acknowledges support from NSF CMMI (award # 2224139). J.S. acknowledges support from ARO (Grant Numbers W911NF-21-2-0169 and W911NF2210048). K.W. and T.T. acknowledge support from the Elemental Strategy Initiative conducted by the MEXT, Japan (Grant Number JPMXP0112101001) and JSPS KAKENHI (Grant Numbers 19H05790, 20H00354 and 21H05233). This work was performed in part at the ASRC NanoFabrication Facility of CUNY in New York. This research used resources of the Center for Functional Nanomaterials (CFN), which is a U.S. Department of Energy Office of Science User Facility, at Brookhaven National Laboratory under Contract No. DE-SC0012704. The authors acknowledge Dr. Suji Park of BNL for help with the operation of the QPress tool.

**Data availability**

Data supporting the findings of this manuscript are available from the corresponding author upon reasonable request.

**Author Contributions**

Z.H., J.S., and D.S. conceived and designed the experiments. Z.H., N.L., B.H.E., E.C., P.Y., M.H., J.S., and D.S. performed the experiments and contributed to the data analysis. K.K. performed the TEM studies. K.W. and T.T. prepared the hBN material. The manuscript was written with input from all authors.

**Competing interests**

The authors declare no competing interest.

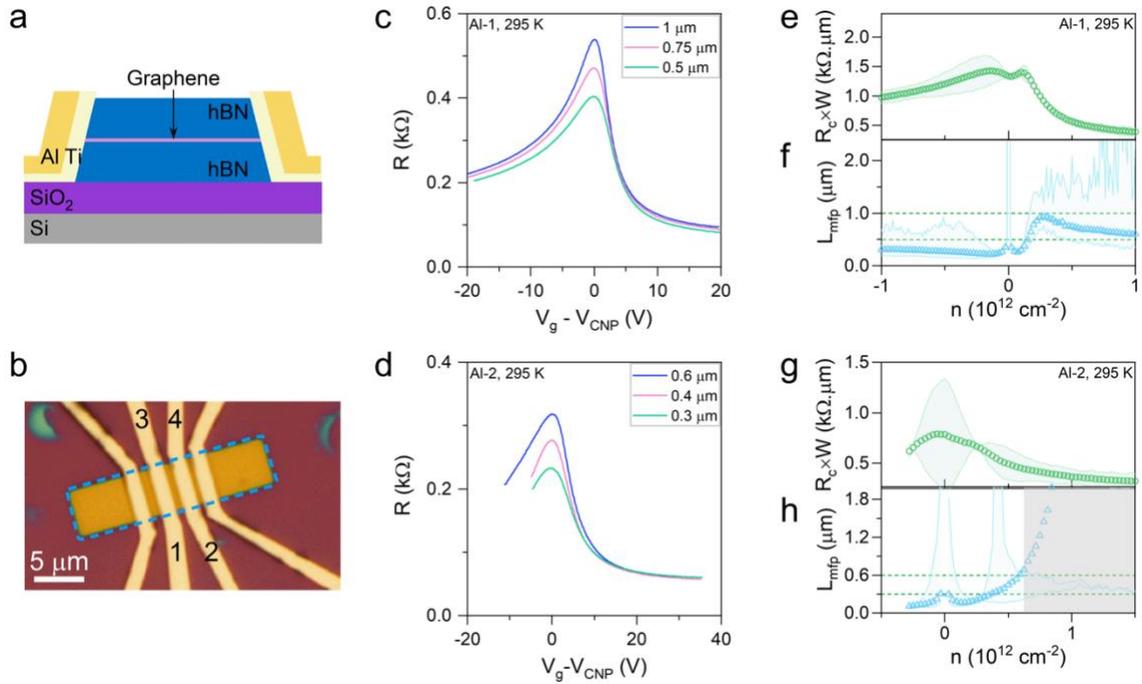

**Figure 1. Analysis of Al-graphene junction at room temperature.** (a) Schematic illustration of the BGB JJ device contacted with Ti/Al. (b) Optical image of the BGB JJ devices from Al-1 sample. The BGB mesa is highlighted with a blue dashed box. Room temperature device resistance as function of gate voltages in (c) Al-1 and (d) Al-2 devices. TLM analysis extracts (e,g) $R_cW$ and (f,h) $L_{mfp}$ from Al-1 and Al-2 devices.

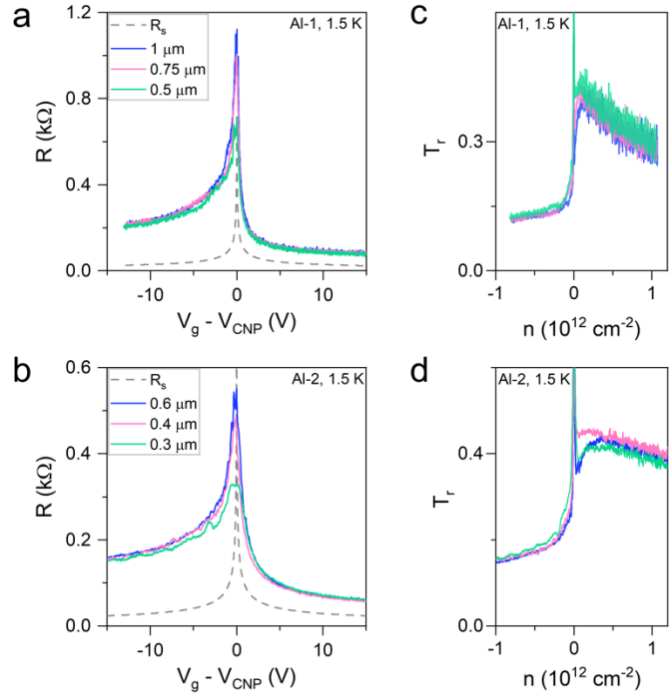

**Figure 2. Low temperature transport analysis.** The graphene resistance as a function of gate voltage measured at 1.5 K for (a) Al-1 and (b) Al-2 devices. $T_r$ calculated from Landauer's formula for (c) Al-1 and (d) Al-2 devices.

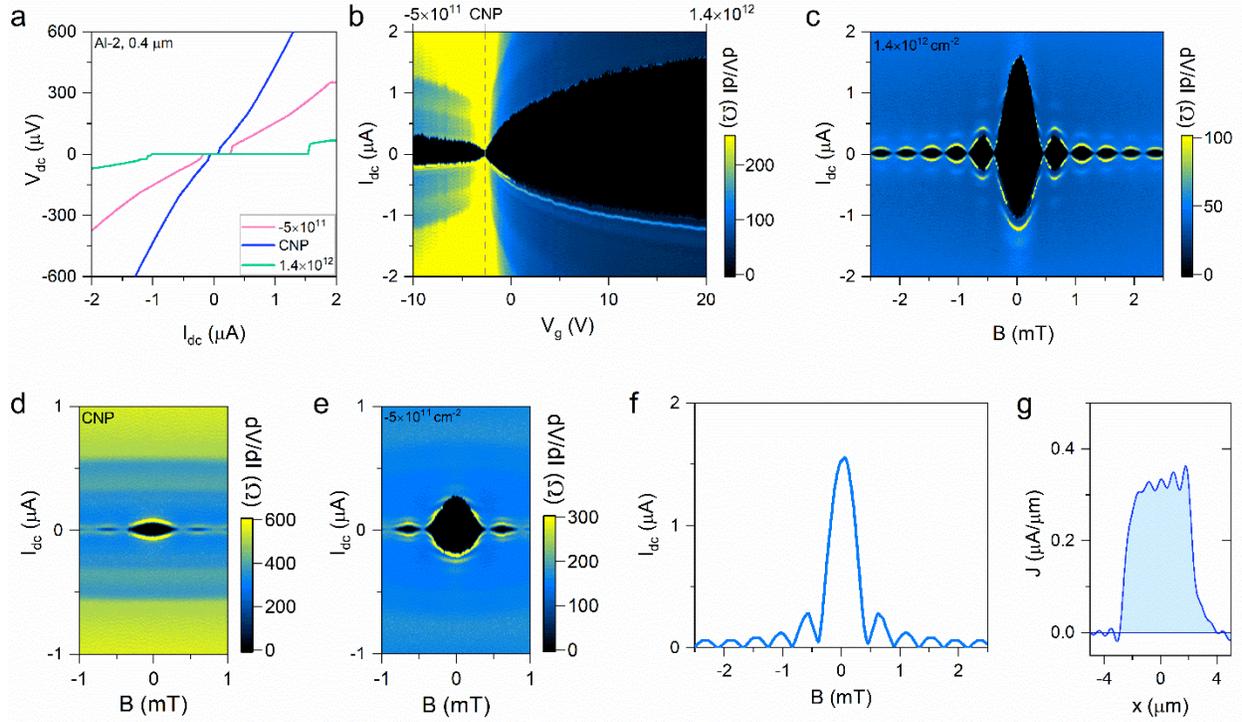

**Figure 3. Proximity-induced superconductivity in graphene.** (a) V-I characteristics measured in the 0.4 μm device at different carrier concentration at 15 mK. (b) Differential resistance as function of bias current and gate voltage in the 0.4 μm BGB JJ device (carrier density legends are in units of cm$^{-2}$. "-" denotes hole carriers). Supercurrent shows Fraunhofer-like interference pattern on the (c) electron, (d) CNP, and (e) hole branch in the presence of a vertical magnetic field. (f) Critical current profile extracted from data in (c). Reconstructed current density distribution from data in (f).

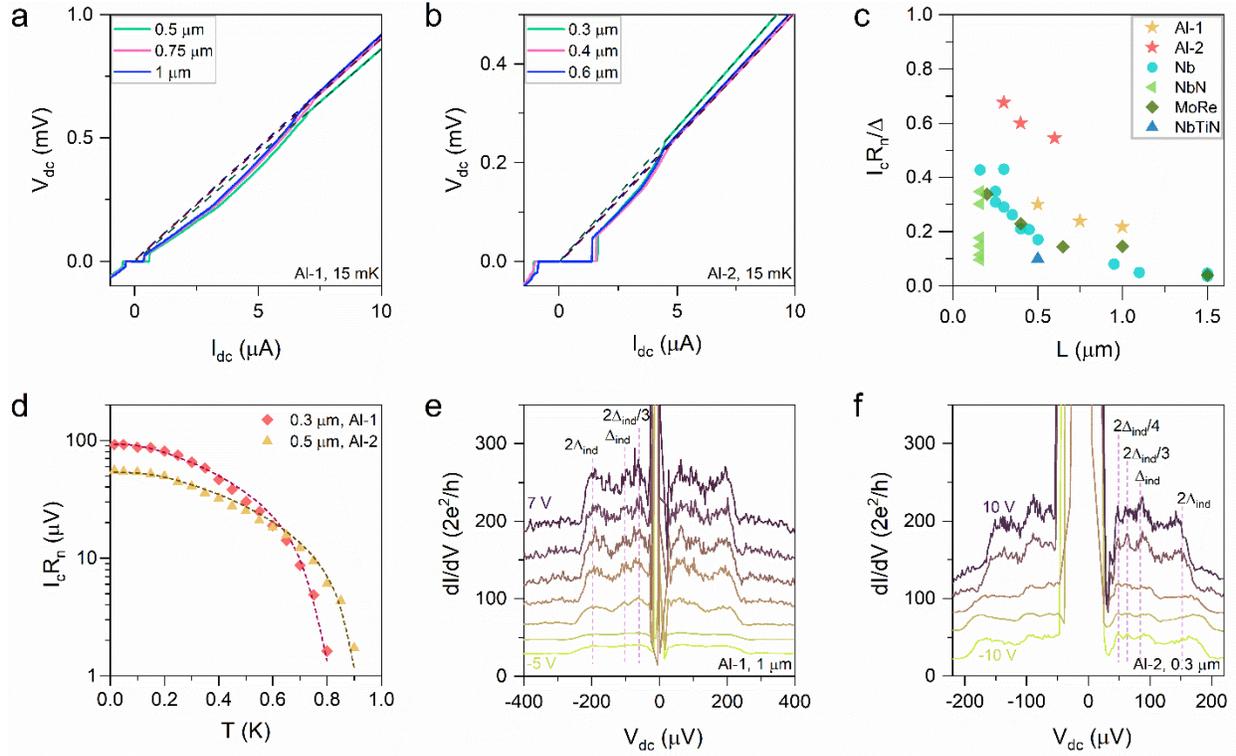

**Figure 4. Junction transparency of Al-graphene devices.** (a)-(b) V-I characteristics of the Al-1 and Al-2 devices. (c) $I_cR_n/\Delta$ plotted as a function of device lengths, comparing our devices with other superconductor edge-contacted BGB devices. (d) Temperature dependent $I_cR_n$ and the fitting for extracting $\alpha$. MAR measures an induced gap of (e) $\Delta_{ind}$=100 μeV in the Al-1 devices and (f) $\Delta_{ind}$=80 μeV in the Al-2 devices. The dashed lines in MAR plots mark the conductance peak position of $2\Delta_{ind}/N$. The dI/dV curves are offset for better illustration.

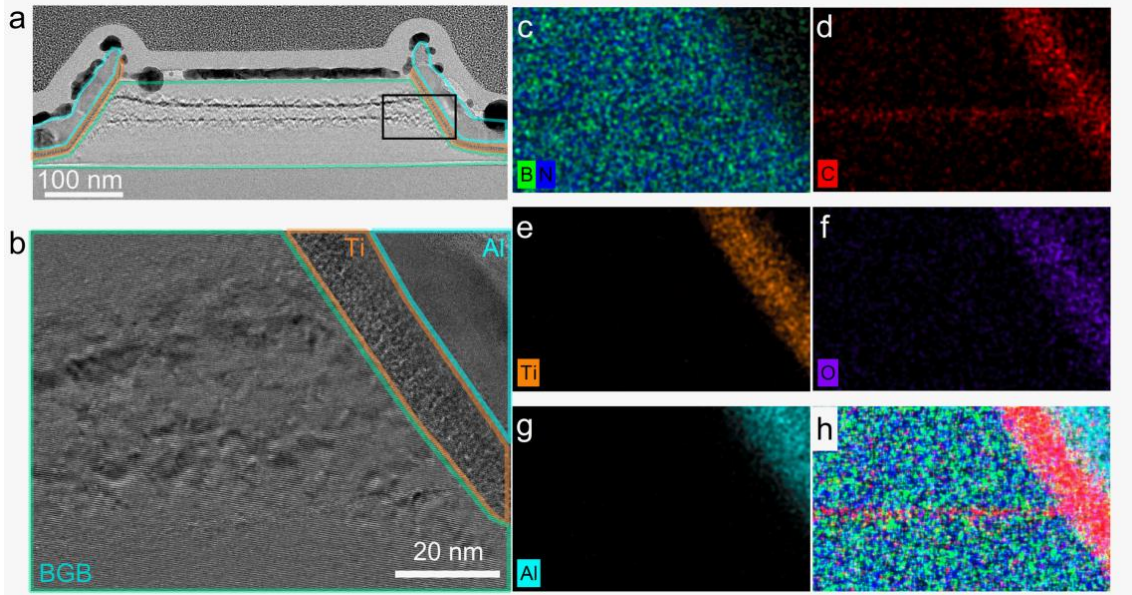

**Figure 5. HRTEM analysis of Al-graphene junction.** (a) Cross-sectional HRTEM image of a BGB JJ with $L_{ch}$=0.5 μm. (b) Zoomed-in HRTEM image at the edge contact region, marked by a solid black box in (a). Elemental composition of the junction shows (c) boron and nitride, (d) C, (e) Ti, (f) O, and (g) Al elements. (h) Overlapping EDS map of all elements.